\documentclass[12pt]{article}

\usepackage{cite}
\usepackage{epsfig}

\catcode`@=11
\def\citer{\@ifnextchar [{\@tempswatrue\@citexr}{\@tempswafalse\@citexr[]}}
 
\def\@citexr[#1]#2{\if@filesw\immediate\write\@auxout{\string\citation{#2}}\fi
  \def\@citea{}\@cite{\@for\@citeb:=#2\do
    {\@citea\def\@citea{--\penalty\@m}\@ifundefined
       {b@\@citeb}{{\bf ?}\@warning
       {Citation `\@citeb' on page \thepage \space undefined}}%
\hbox{\csname b@\@citeb\endcsname}}}{#1}}
\catcode`@=12

\def\refeq#1{\mbox{eq.~(\ref{#1})}}

\def\reffi#1{\mbox{Fig.~\ref{#1}}}

\def\citere#1{\mbox{Ref.~\cite{#1}}}
\def\citeres#1{\mbox{Refs.~\cite{#1}}}

\newcommand{\MstL}{M_{\tilde{t}_L}}
\newcommand{\MstR}{M_{\tilde{t}_R}}

\newcommand{\At}{A_t}

\newcommand{\Xt}{X_t}

\newcommand{\ms}{M_S}
\newcommand{\msusy}{M_{\mathrm{SUSY}}}
\newcommand{\msq}{m_{\tilde{q}}}

\newcommand{\msbar}{$\overline{\rm{MS}}$}

\newcommand{\oaas}{{\cal O}(\alpha\alpha_s)}

\newcommand{\cp}{{\cal CP}}
\newcommand{\wz}{\sqrt{2}}
\newcommand{\edz}{\frac{1}{2}}

\newcommand{\twol}{two-loop}
\newcommand{\onel}{one-loop}

\newcommand{\fh}{{\em FeynHiggs}}

\newcommand{\subh}{{\em subhpole}}

\newcommand{\MZ}{M_Z}
\newcommand{\MA}{M_A}
\newcommand{\mh}{m_h}
\newcommand{\mhmax}{m_h^{\rm max}}

\newcommand{\mt}{m_{t}}
\newcommand{\mtms}{\overline{m}_t}

\newcommand{\Stop}{\tilde{t}}
\newcommand{\StopL}{\tilde{t}_L}
\newcommand{\StopR}{\tilde{t}_R}

\newcommand{\tsf}{\theta\kern-.20em_{\tilde{f}}}
\newcommand{\tsfp}{\theta\kern-.20em_{\tilde{f}\prime}}
\newcommand{\tsq}{\theta\kern-.15em_{\tilde{q}}}
\newcommand{\sw}{s_W}

\newcommand{\KL}{\left(}
\newcommand{\KR}{\right)}

\newcommand{\VL}{\left( \begin{array}{c}}
\newcommand{\VR}{\end{array} \right)}
\newcommand{\ML}{\left( \begin{array}{cc}}
\newcommand{\MLd}{\left( \begin{array}{ccc}}
\newcommand{\MLv}{\left( \begin{array}{cccc}}
\newcommand{\MR}{\end{array} \right)}

\newcommand{\tb}{\tan \beta}

\newcommand{\CTb}{\cot \beta\hspace{1mm}}

\newcommand{\CZb}{\cos 2\beta\hspace{1mm}}

\newcommand{\tev}{\,\, \mathrm{TeV}}
\newcommand{\gev}{\,\, \mathrm{GeV}}
\newcommand{\mev}{\,\, \mathrm{MeV}}

\newcommand{\BC}{\begin{center}}
\newcommand{\EC}{\end{center}}
\newcommand{\BE}{\begin{equation}}
\newcommand{\EE}{\end{equation}}
\newcommand{\BEA}{\begin{eqnarray}}
\newcommand{\BEAnn}{\begin{eqnarray*}}
\newcommand{\EEA}{\end{eqnarray}}
\newcommand{\EEAnn}{\end{eqnarray*}}
\newcommand{\non}{\nonumber}
\newcommand{\id}{{\rm 1\kern-.12em
\rule{0.3pt}{1.5ex}\raisebox{0.0ex}{\rule{0.1em}{0.3pt}}}}
\newcommand{\lsim}
{\;\raisebox{-.3em}{$\stackrel{\displaystyle <}{\sim}$}\;}

\newcommand{\gf}{G_F}

\def\al{\alpha}

\def\als{\alpha_s}

\def\De{\Delta}

\newcommand{\lmtmsms}{\KL\frac{\mtms^2}{\ms^2}\KR}

\begin{document}
\thispagestyle{empty}

\def\thefootnote{\fnsymbol{footnote}}

\begin{flushright}
DESY 99-120\\
KA-TP-12-1999\\
hep-ph/9909540\\
\end{flushright}

\vspace{1cm}

\begin{center}

{\large\sc {\bf Constraints on $\tb$ in the MSSM from the Upper Bound}}

\vspace*{0.4cm} 

{\large\sc {\bf on the Mass of the Lightest Higgs boson}}

\vspace{1cm}

{\sc 
S.~Heinemeyer$^{1}$%
\footnote{email: Sven.Heinemeyer@desy.de}%
, W.~Hollik$^{2}$%
\footnote{email: Wolfgang.Hollik@physik.uni-karlsruhe.de}%
and G.~Weiglein$^{3}$%
\footnote{email: Georg.Weiglein@cern.ch}
}

\vspace*{1cm}

{\sl
$^1$ DESY Theorie, Notkestr. 85, D--22603 Hamburg, Germany

\vspace*{0.4cm}

$^2$ Institut f\"ur Theoretische Physik, Universit\"at Karlsruhe, \\
D--76128 Karlsruhe, Germany

\vspace*{0.4cm}

$^3$ Theoretical Physics Division, CERN, CH-1211 Geneva 23, Switzerland
}

\end{center}

\vspace*{2cm}

\begin{abstract}
We investigate the possibilities for constraining $\tb$ within the MSSM
by combining the theoretical result for the upper bound on the lightest 
Higgs-boson mass as a function of $\tb$ with the informations from the
direct experimental search for this particle. We discuss the commonly used
``benchmark'' scenario, in which the parameter values $\mt = 175$~GeV
and $\msusy = 1$~TeV are chosen, and analyze in detail the effects
of varying the other SUSY parameters. We furthermore study the impact of
the new diagrammatic two-loop result for $\mh$, which leads to an
increase of the upper bound on $\mh$ by several GeV, on present and future
constraints on $\tb$. We suggest a slight generalization of the
``benchmark'' scenario, such that the scenario contains the maximal
possible values for $\mh(\tb)$ within the MSSM for 
fixed $\mt$ and $\msusy$.
The implications of
allowing values for $\mt$, $\msusy$ beyond the ``benchmark'' scenario
are also discussed.

\end{abstract}

\vspace{.5cm}


\def\thefootnote{\arabic{footnote}}
\setcounter{page}{0}
\setcounter{footnote}{0}

\newpage


\section{Theoretical basis}

Within the MSSM the masses of the $\cp$-even neutral Higgs bosons are
calculable in terms of the other MSSM parameters. The mass of the
lightest Higgs boson, $\mh$, has been of particular interest: \onel\
calculations~\cite{mhiggs1l,mhiggsf1l} have been supplemented in the
last years with the leading \twol\ corrections, performed in the
renormalization group (RG)
approach~\cite{mhiggsRG1,mhiggsRG1a,mhiggsRG1b,mhiggsRG2}, in the effective
potential approach~\cite{hoanghempfling,zhang} and most recently in
the Feynman-diagrammatic (FD)
approach~\cite{mhiggsletter,mhiggslong}. These calculations predict an
upper bound on $\mh$ of about $\mh \lsim 135 \gev$.

For the numerical evaluations in this paper we made use of the
Fortran code \subh, corresponding to the RG calculation~\cite{mhiggsRG1b}, 
and of the
program \fh~\cite{feynhiggs}, corresponding to the recent result of 
our FD calculation.

\bigskip
In order to fix our notations, we list the conventions for the input
from the scalar top sector of the MSSM:
the mass matrix in the basis of the current eigenstates $\StopL$ and
$\StopR$ is given by
\BE
\label{stopmassmatrix}
{\cal M}^2_{\Stop} =
  \ML \MstL^2 + \mt^2 + \CZb (\edz - \frac{2}{3} \sw^2) \MZ^2 &
      \mt \Xt \\
      \mt \Xt &
      \MstR^2 + \mt^2 + \frac{2}{3} \CZb \sw^2 \MZ^2 
  \MR,
\EE
where 
\BE
\mt \Xt = \mt (A_t - \mu \CTb)~.
\label{eq:mtlr}
\EE
For the numerical evaluation, a common choice is
\BE
\MstL = \MstR =: \msusy ;
\EE
this has been shown to yield upper values for $\mh$ which comprise
also the case where $\MstL \neq \MstR$,
when $\msusy$ is identified with the heavier one~\cite{mhiggslong}.
We furthermore use the short-hand notation
\BE
\ms^2 := \msusy^2 + \mt^2~.
\EE

\begin{figure}[ht!]
\vspace{1em}
\begin{center}
\mbox{
\psfig{figure=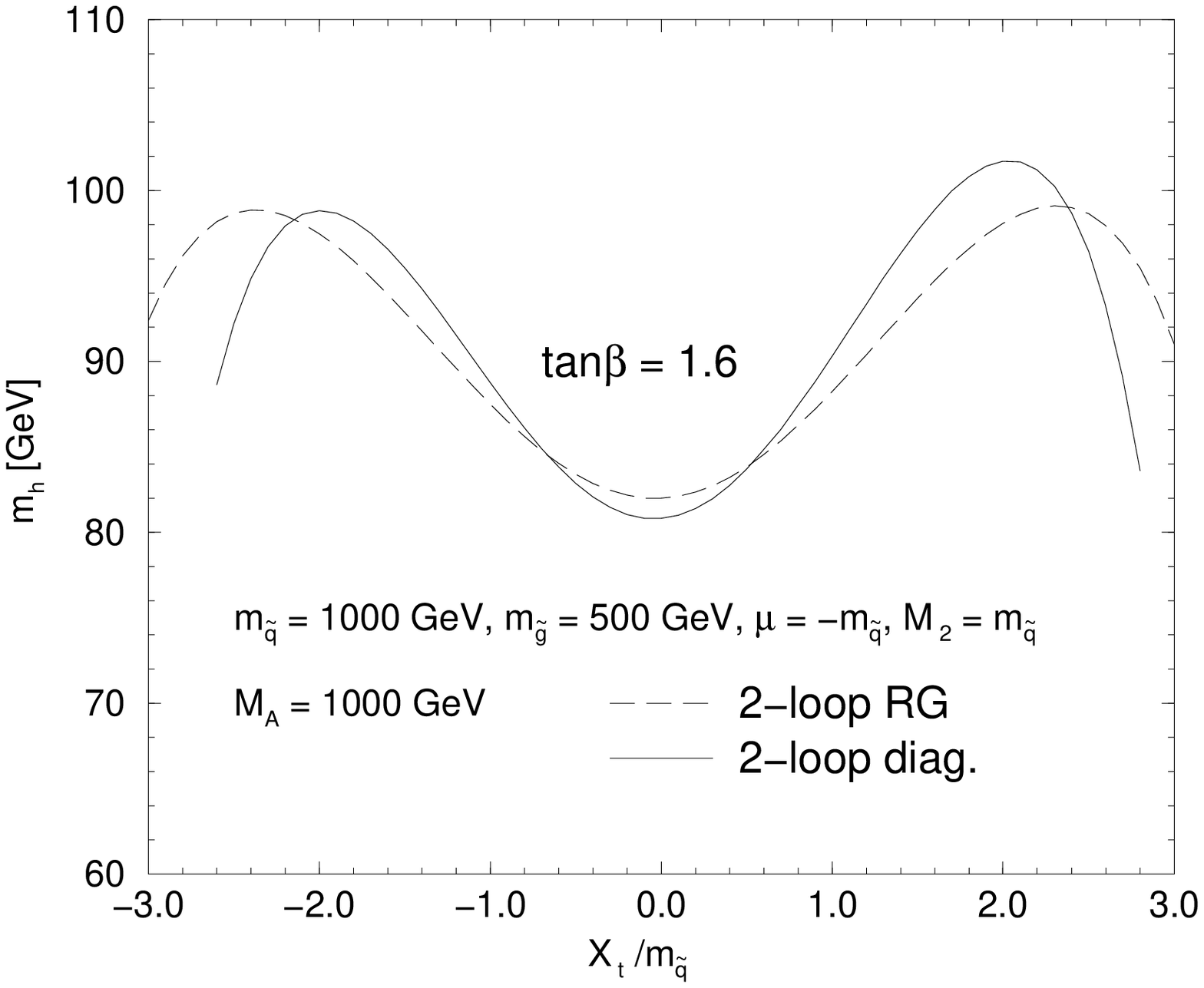,width=11.5cm}}
\end{center}
\caption[]{\it $\mh$ is shown as a function of $\Xt/\msq$ for $\tb = 1.6$
evaluated in the Feynman-diagrammatic (program \fh) and in the 
renormalization group (program \subh) approach, where $\msq \equiv \msusy$. 
The maximal value of $\mh$ is obtained for 
$\Xt \approx 2 \, \msq$ in the FD approach and $\Xt \approx 2.4 \, \msq$
in the RG approach.
}
\label{fig:fdrg}
\end{figure}

While the case $\Xt = 0$ is labelled as `no-mixing', it is customary to
to assign `maximal-mixing' to the value of $\Xt$ for which the
the mass of the lightest Higgs boson is maximal. As can be seen in
\reffi{fig:fdrg}, where $\mh$ is shown as a function of $\Xt/\msusy$
within the FD and the RG approach, the `maximal-mixing' case corresponds
to $\Xt \approx 2\, \msusy$ in the FD approach, while it corresponds to
$\Xt = \sqrt{6}\, \msusy \approx 2.4 \, \msusy$ in the RG approach. It
should be noted in this context that, due to the different
renormalization schemes utilized in the FD and the RG approach, the
(scheme-dependent) parameters $\Xt$ and $\msusy$ have a different meaning
in the two approaches, which has to be taken into account when comparing
the corresponding results. While the resulting shift in $\msusy$ turns
out to be small, sizable differences occur between the numerical values of 
$\Xt$ in the two schemes, see \citeres{mhiggslong,bse}.

\bigskip
The main differences between the RG and the FD calculation have been
investigated in \citeres{bse,mhiggslle}. They arise at the \twol\ level.
The dominant \twol\ contribution of $\oaas$ to $\mh^2$ in the FD
approach reads:
\BEA 
\label{mh2twolooptop}
\De\mh^{2,\al\als} &=& \De m_{h,\rm log}^{2,\al\als} 
                     + \De m_{h,\rm non-log}^{2,\al\als}~, \\
\De m_{h,\rm log}^{2,\al\als} &=&
    - \frac{\gf\wz}{\pi^2} \frac{\als}{\pi}\; \mtms^4
      \Biggl[ 3 \log^2\lmtmsms + 2 \log\lmtmsms 
     - 3 \frac{\Xt^2}{\ms^2} \log\lmtmsms \Biggr]~, \\
\De m_{h,\rm non-log}^{2,\al\als} &=& 
    - \frac{\gf\wz}{\pi^2} \frac{\als}{\pi}\; \mtms^4
      \Biggl[ 4 -6 \frac{\Xt}{\ms} 
     - 8 \frac{\Xt^2}{\ms^2} 
     +\frac{17}{12} \frac{\Xt^4}{\ms^4} \Biggr]~;
\EEA
therein $\mtms$ denotes the running top-quark mass
\BE
\mtms \equiv \mtms(\mt) \approx \frac{\mt}{1 + \frac{4}{3\,\pi} \als(\mt)}~.
\label{mtrun}
\EE

By transforming the FD result into the \msbar\ scheme,
it has been shown analytically that the RG and the FD approach agree
in the logarithmic terms~\cite{bse}. The non-logarithmic terms
$\De m_{h,\rm non-log}^{2,\al\als}$, however, are genuine
two-loop terms, obtained by explicit diagrammatic
calculation~\cite{bse,mhiggslle}.  
In the maximal-mixing scenario,
these terms can enhance the lightest Higgs-boson mass by up to 
$5 \gev$ (see also the discussion of \reffi{fig:rgv} and the
corresponding footnote.)

The new two-loop terms obtained within the FD approach lead to a
reduction of the theoretical uncertainty of the Higgs-mass prediction
due to unknown higher-order corrections (see \citere{bse} for a
discussion). Another source of theoretical uncertainty is related to the
experimental errors of the input parameters, such as $\mt$. In 
the case of the SUSY parameters, direct experimental information is
lacking completely. For this reason it is convenient to discuss
specific scenarios, where certain values of the parameters are assumed.


\section{The benchmark scenario}

In recent years it has become customary to discuss the restrictions on 
$\tb$ from the search for the lightest Higgs boson within the so-called
``benchmark'' scenario, which is specified by the parameter values
\BE
\mt = 175 \gev, \qquad \msusy = 1 \tev, 
\label{eq:bench}
\EE
where $\msusy$ denotes the common 
soft SUSY breaking scale for all sfermions (see e.g.
\citeres{Higgsgroup,aleph,l3,delphi,opalprep}
for recent analyses within this framework). According to 
\citeres{Higgsgroup,opalprep,hzha,tomj}, the other SUSY parameters within
the benchmark scenario are chosen as
\BEA
\mu &=& -100 \gev \non \\
M_2 &=& 1630 \gev \non \\
\MA &\leq& 500 \gev \non \\
\At &=& 0 \quad \mbox{(``no mixing'')} \non \\
\At &=& \sqrt{6}\, \msusy \quad \mbox{(``maximal mixing'')} ,
\label{benchmarkdef}
\EEA
where $\mu$ is the Higgs mixing parameter, $M_2$ denotes the
soft SUSY breaking parameter in the gaugino sector, and $\MA$ is 
the $\cp$-odd Higgs-boson mass. The maximal possible Higgs-boson mass 
as a function of $\tb$ within this scenario is obtained for 
$\At = \sqrt{6} \, \msusy$ and $\MA = 500$~GeV. Exclusion limits on $\tb$ 
within this scenario follow by combining the information from the
theoretical upper bound in the $\tb$--$\mh$ plane with the direct search 
results for the lightest Higgs boson.

\bigskip
The tree-level value for $\mh$ within the MSSM is determined by 
$\MA$, $\tb$ and the $Z$-boson mass $\MZ$. 
Beyond the tree-level, 
the main correction to $\mh$ stems from the $t$--$\Stop$-sector. Thus, the
most important parameters for the corrections to $\mh$ are $\mt$,
$\msusy$ and $\Xt$. 

\begin{figure}[ht!]
\vspace{1em}
\begin{center}
\mbox{
\psfig{figure=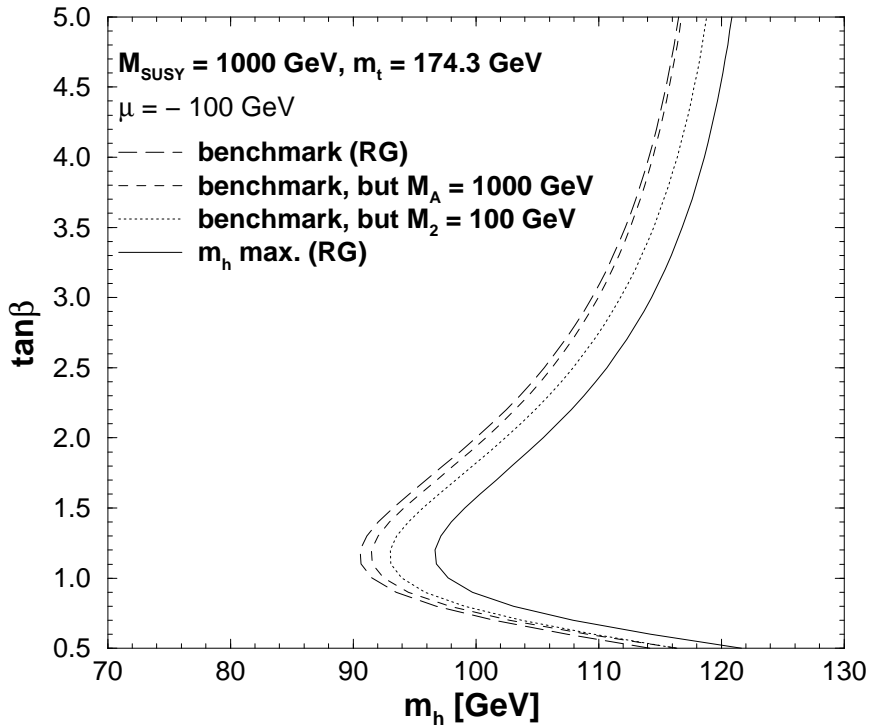,width=11.5cm}}
\end{center}
\caption[]{\it $\mh$ is shown as a function of $\tb$, evaluated in the
RG approach.
The left (long-dashed) curve displays the benchmark scenario. For the 
dotted (dashed) curves one deviation from the benchmark scenario, 
$M_2 = 100 \gev$
($\MA = 1000 \gev$), is taken into account. The solid curve displays the
maximal possible $\mh$ value for $\mt = 174.3 \gev$ and $\msusy = 1 \tev$.
}
\label{fig:bmdev}
\end{figure}

Since the benchmark scenario relies on specifying the two parameters
$\mt = 175$~GeV and $\msusy = 1$~TeV, it is of interest to investigate
whether the other inputs in the benchmark scenario are allowed to vary
in such a way that the maximal possible value for $\mh$, once $\mt$ and
$\msusy$ are fixed, is contained in this scenario. This is however not
the case:

\begin{itemize}
\item
Compared to the ``benchmark'' value of $M_2 = 1630 \gev$, the value of
$\mh$ is enhanced by about $2.5 \gev$ (depending slightly 
on the value of $\tb$) by choosing a small value for 
$M_2$, e.g.\ $M_2 = 100 \gev$ (see \citere{mhiggslong}, where
a scan over the MSSM parameter space has been performed showing that 
the maximal values for $\mh$ are obtained for small values of 
$M_2$ and $|\mu|$).

\item
While in the benchmark scenario only $\MA$ values up to 500~GeV are
considered, higher $\MA$ values lead to an increase of $\mh$. For $\MA =
1000 \gev$, $\mh$ is enhanced by up to $1.5 \gev$.

\item 
While within the benchmark scenario ``maximal mixing'' is defined as 
\BE
\At = \Xt + \mu \CTb = \sqrt{6}\,\msusy~,
\EE
the maximal Higgs-boson masses are in fact obtained (in the RG approach)
for 
\BE
\Xt = \sqrt{6}\,\msusy~~({\rm RG})~.
\EE
This changes $\mh$ by ${\cal O}(300 \mev)$ for $\tb = {\cal O}(1.6)$
and $\mu = -100 \gev$. 
As mentioned above, in the FD calculation one has to take 
\BE
\Xt = 2\,\msusy~~({\rm FD})
\EE
for maximal mixing.%
\footnote{
As already explained in Sect.\ 1,
the different values for $\Xt$ yielding the maximal $\mh$ values in
the FD and in the RG approach reflect the fact that this
(unobservable) parameter has a different meaning in both approaches
due to the different renormalization schemes employed. This has been
analyzed in detail in \citere{bse}. Thus using different $\Xt$
values in the FD and the RG calculation takes this scheme difference into
account and individually maximizes the $\mh$ values, see \reffi{fig:fdrg}.
}

\item
In the benchmark scenario, according to the implementation in the 
HZHA event generator~\cite{hzha}, the running top-quark mass has been
defined by including corrections up to ${\cal O}(\als^2)$. Compared to
the definition~(\ref{mtrun}), which includes only corrections up to
${\cal O}(\als)$, this leads to a reduction of the running top-quark
mass by about $2 \gev$. From the point of view of a perturbative
calculation up to ${\cal O}(\alpha\als)$ it is however not clear 
whether corrections of ${\cal O}(\als^2)$ in the running top-quark mass,
which is inserted into an expression of ${\cal O}(\alpha)$, will in fact 
lead to an improved result. On the contrary, as a matter of consistency 
of the perturbative evaluation it
appears to be even favorable to restrict the running top-quark mass to
its ${\cal O}(\als)$ expression~(\ref{mtrun}). Adopting this more
conservative choice leads to an increase of $\mh$ by up to $1.5 \gev$.
\end{itemize}
All four effects shift the Higgs-boson mass to higher values.
For the analyses below we will use the current experimental value for
the top-quark mass, $\mt = 174.3 \gev$~\cite{top}, 
i.e.\ we consider the benchmark
scenario with $\mt = 174.3 \gev$ and $\msusy = 1$~TeV.
Two of the effects discussed above are displayed in \reffi{fig:bmdev}, 
where also the maximal values for $\mh$ according to the discussion
above ($\mhmax$-scenario: $M_2 = 100$~GeV, $\MA = 1000$~GeV, 
$\Xt = \sqrt{6}\,\msusy$ (RG), $\Xt = 2\,\msusy$ (FD),
$\mtms$ as defined in \refeq{mtrun}) obtained in the 
RG approach with $\mt = 174.3 \gev$ and $\msusy = 1 \tev$ are
displayed. Comparing the $\mhmax$-scenario with the benchmark scenario,
the values for $\mh$ are higher by about $5 \gev$.

\begin{figure}[ht!]
\vspace{1em}
\begin{center}
\mbox{
\psfig{figure=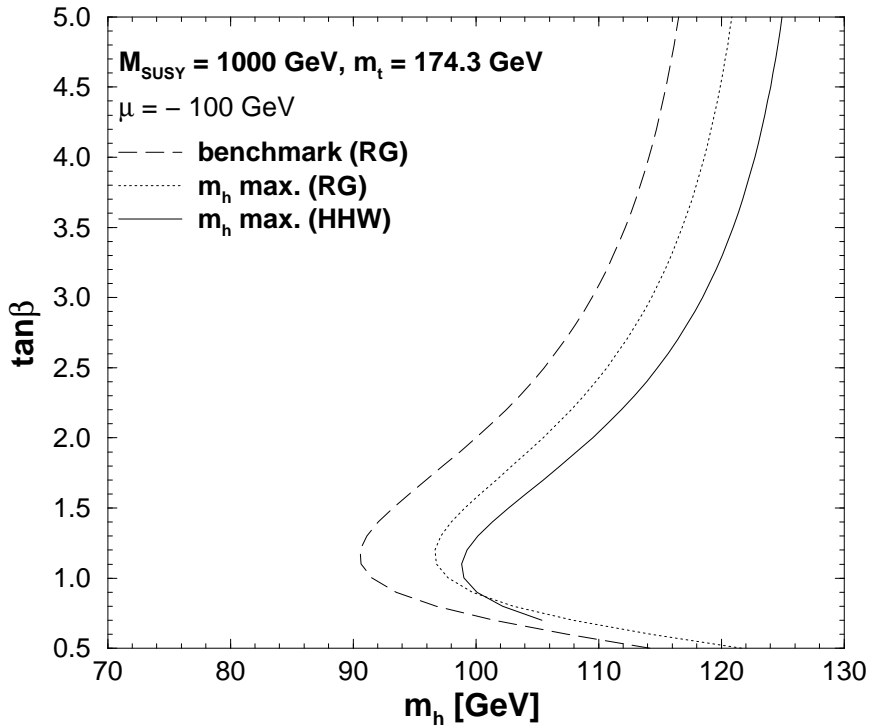,width=11.5cm}}
\end{center}
\caption[]{\it $\mh$ is shown as a function of $\tb$.
The dashed curve displays the benchmark scenario. 
The dotted curve shows the $\mhmax$-RG scenario (program \subh), 
while the solid curve represents the $\mhmax$-FD scenario (HHW, program \fh).
}
\label{fig:rgv}
\end{figure}

\bigskip
So far we have only discussed the increase in the maximal value of the
Higgs-boson mass which is obtained using the slight generalization of the 
benchmark scenario discussed above. Now we also take into account the
impact of the new FD two-loop result for $\mh$, which contains
previously unknown non-logarithmic two-loop terms. The corresponding
result in the $\tb$--$\mh$ plane (program \fh) is shown in \reffi{fig:rgv} 
in comparison with the benchmark scenario and the $\mhmax$-RG scenario
(program \subh).
The maximal value for $\mh$ within the FD result is higher by up to $4
\gev$ compared to the $\mhmax$-RG scenario%
\footnote{
In \citere{bse} it has recently been shown that (in the 
leading $\mt^4$ corrections to $\mh$) a large part of the genuine
\twol\ corrections included in the FD calculation can be absorbed by
an appropriate scale choice of the running top-quark mass into an
effective \onel\ result. Modifying the RG result by using this scale
choice for the running top-quark mass would
lead to an increase of the RG curve in \reffi{fig:rgv} by up to 3~GeV,
leaving only a difference of 1-2~GeV between the FD and the RG
result. 
}
 and by up to 9~GeV compared
to the benchmark scenario.

The increase in the maximal value for $\mh$ by about $4$~GeV from the
new FD result and by further 5~GeV if the benchmark scenario is
slightly generalized has a significant effect on exclusion limits for $\tb$
derived from the Higgs-boson search. Employing the benchmark scenario
and the RG result, an excluded $\tb$~range 
already appears for an experimental bound on $\mh$ of slightly above 90~GeV,
see \reffi{fig:bmdev}. However, taking into account the above sources 
for an increase in the maximal value for $\mh$ the current data 
(summer '99, see e.g.\ \citere{aqtalk}) from the
Higgs-boson search hardly allow any $\tb$ exclusion yet, see \reffi{fig:rgv}. 
Concerning the assumed $\mh$ limit obtained at the end of LEP2, the accessible
$\tb$ region is largely reduced from the $\mhmax$-RG to the $\mhmax$-FD
calculation.


\section{ Constraints on $\tb$ ``beyond the benchmark''}

\begin{figure}[ht!]
\vspace{2em}
\begin{center}
\mbox{
\psfig{figure=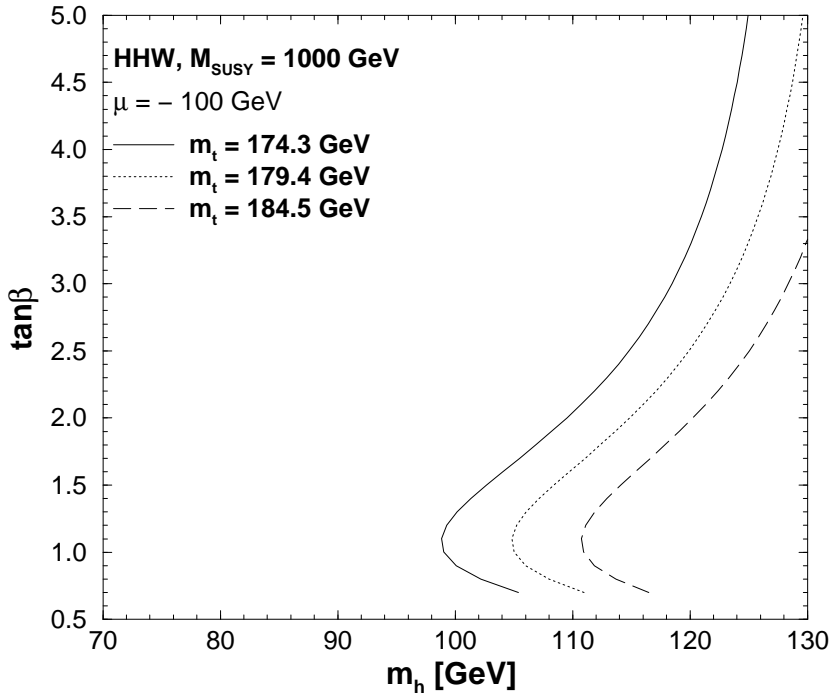,width=11cm}}
\end{center}
\caption[]{\it $\mh$ is shown as a function of $\tb$, evaluated in the
FD approach. We give the results for three different values of the
top-quark mass, $\mt = 174.3, 179.4, 184.5 \gev$.
}
\label{fig:mhmt}
\end{figure}

Since the dominant radiative corrections to the lightest Higgs-boson
mass are proportional to $\mt^4$, the theoretical prediction for $\mh$
depends sensitively on the precise value of the top-quark mass.
The experimental uncertainty in the top-quark mass of currently 
$\De\mt = 5.1 \gev$~\cite{top}
thus has a strong effect on the prediction for the upper bound on $\mh$,
where larger values of $\mt$ give rise
to larger values of $\mh$. An increase in $\mt$ by $\De\mt = 5.1 \gev$
leads to an increase in $\mh$ of up to 6~GeV, as shown in
\reffi{fig:mhmt}, where also the effect of increasing $\mt$ by two
standard deviations is displayed.

Besides the top-quark mass, the other main entry of the benchmark
scenario is the choice
$\msusy = 1 \tev$. Similarly to the case of $\mt$, allowing for higher
values of $\msusy$ leads to higher values of $\mh$. Since $\msusy$
enters only logarithmically in the prediction for $\mh$, the dependence
on it is more moderate. An increase of $\msusy$ from 1~TeV to 2~TeV
enhances $\mh$ by up to 4~GeV (depending on $\tb$).

\begin{figure}[ht!]
\vspace{1em}
\begin{center}
\mbox{
\psfig{figure=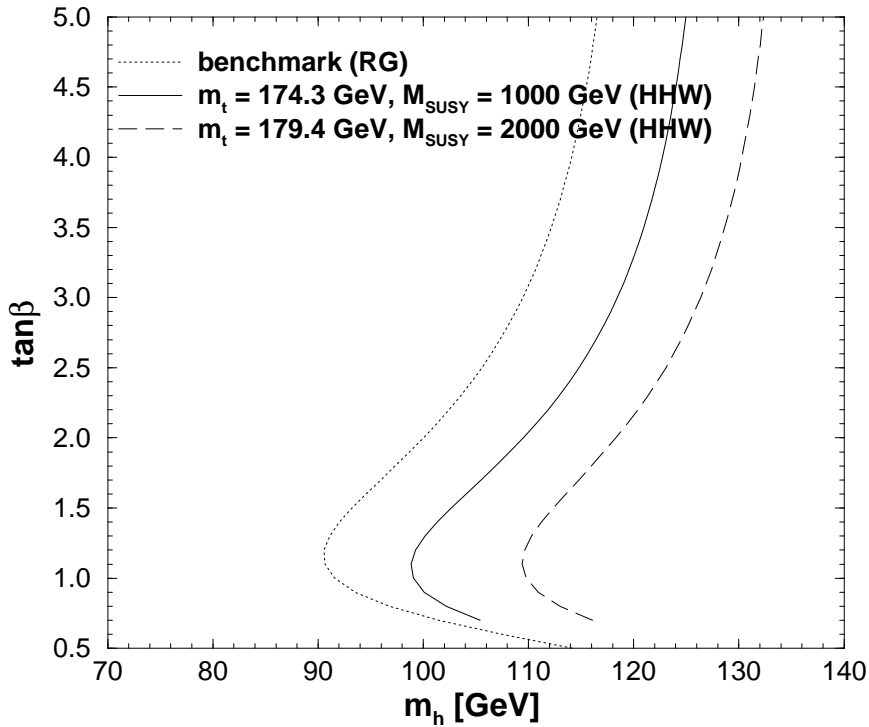,width=11.5cm}}
\end{center}
\caption[]{\it $\mh$ is shown as a function of $\tb$.
The dotted curve displays the benchmark scenario in the RG approach,
which has been used for phenomenological analyses up to now. 
The solid curve displays the $\mhmax$-FD scenario, while the dashed curve 
corresponds to the ``increased $\mh$'' scenario with $\mt = 179.4 \gev$ and 
$\msusy = 2000 \gev$.
}
\label{fig:worstcase}
\end{figure}

Allowing values of $\mt$ one or even two standard deviations above the
current experimental central value and increasing also the input value of
$\msusy$ clearly has a large effect on possible $\tb$ constraints.
In \reffi{fig:worstcase} we show an ``increased $\mh$'' scenario,
where $\mt = 179.4 \gev$ has been chosen, i.e.\ one standard deviation
above the current experimental value, and $\msusy = 2000 \gev$ is taken.
It is compared with the benchmark scenario in the RG calculation and 
with the $\mhmax$-FD scenario. In the ``increased $\mh$'' scenario 
exclusion of a $\tb$ range would become possible only with a limit on
$\mh$ of more than about $110 \gev$.

In this context one should keep in mind that the benchmark scenario
contains not only an assumption about the SUSY parameters but also about
the actual model which is tested, namely a SUSY model with a minimal
Higgs sector that does not contain $\cp$-violating phases. 
The upper bound on $\mh$, however, stays the same also with complex
parameters~\cite{cphiggs}. 
Extensions of
the Higgs sector by additional particle representations
can shift the upper bound on the mass of the lightest Higgs boson up to
values of about 200~GeV~\cite{espi}.


\section{Conclusions}

We have investigated the upper bound on the mass of the lightest 
$\cp$-even Higgs boson in the MSSM, depending on $\tb$. In order to
discuss possible exclusion limits on $\tb$ from the direct Higgs-boson
search, it is useful to consider definite scenarios with specific
assumptions on the relevant input parameters and on the structure of the
considered model. Constraints on $\tb$ derived within such a framework
are of course to be understood under the assumptions defining the
investigated scenario.

In this spirit in particular the ``benchmark'' scenario has been widely
used, in which $\mt = 175$~GeV and $\msusy = 1$~TeV are chosen. In this
note we have analyzed the influence of variations in the other parameters 
entering the prediction for $\mh$ and we have shown that the settings used 
for those parameters within the benchmark scenario do not cover the maximal
possible value of $\mh$ for $\mt = 175$~GeV and $\msusy = 1$~TeV. We
thus suggest a slight generalization of the definition of the benchmark
scenario, where more general values of $M_2$ and $M_A$ are allowed, a
more conservative expression for the running top-quark mass is taken,
and the case of maximal mixing in the scalar top sector is defined such
that it corresponds to the maximal $\mh$ value. 
Compared to the definition of the benchmark scenario used so far, the
generalization suggested here leads to a shift in the upper bound of
$\mh$ of about 5~GeV.

Independently of the precise definition of the benchmark scenario, we
have furthermore analyzed the impact of taking into account the new 
diagrammatic two-loop result (program \fh) for the mass of the lightest 
Higgs boson, which contains in particular genuine non-logarithmic two-loop 
contributions that are not present in the previous result obtained by 
renormalization group methods (program \subh). The maximal value for 
$\mh$ obtained with \fh\ is higher by about 4~GeV than the maximal value
calculated with \subh. This leads to a significant reduction of the 
$\tb$ region accessible at LEP2.

Going beyond the benchmark scenario, we have also discussed an 
``increased $\mh$'' scenario, where $\mt$ is chosen to be one standard
deviation 
above the current experimental central value and $\msusy = 2$~TeV. In this
scenario no values of $\tb$ can be excluded as long as the limit on
$\mh$ is lower than about $110 \gev$.


\section*{Acknowledgements}
We thank M.~Carena, K.~Desch, E.~Gross, P.~Janot, T.~Junk, A.~Quadt
and C.~Wagner for valuable discussions.




\end{document}